\documentclass[prb,twocolumn,showpacs,floatfix,preprintnumbers,amsmath,amssymb,superscriptaddress]{revtex4-1}
\usepackage{graphicx}
\usepackage{color}\usepackage{float}

\begin{document}


\title{Superconducting Order Parameter and Bosonic Mode in Hydrogen-Substituted NdFeAsO$_{0.6}$H$_{0.36}$ Revealed by Multiple Andreev Reflection Spectroscopy}


\author{T. E. Kuzmicheva}
\email[]{kute@sci.lebedev.ru}

\affiliation{Lebedev Physical Institute, Russian Academy of Sciences, 119991 Moscow, Russia}

\author{S. A. Kuzmichev}
\affiliation{Lomonosov Moscow State University, Faculty of Physics, 119991 Moscow, Russia}
\affiliation{Lebedev Physical Institute, Russian Academy of Sciences, 119991 Moscow, Russia}

\author{N. D. Zhigadlo}
\affiliation{Department of Chemistry and Biochemistry, University of Bern,
CH-3012 Bern, Switzerland}
\affiliation{Laboratory for Solid State Physics, ETH Zurich,
CH-8093 Zurich, Switzerland}
\affiliation{CrystMat Company, CH-8046 Zurich, Switzerland}


\date{\today}

\begin{abstract}
Using intrinsic multiple Andreev reflections effect (IMARE) spectroscopy, we studied ballistic superconductor - normal metal - superconductor (SnS) contacts in layered oxypnictide superconductors NdFeAsO$_{0.6}$H$_{0.36}$ with critical temperatures $T_c = 45-48$\,K. We directly determined the magnitude of two bulk superconducting order parameters, the large gap $\Delta_L \approx 10.4$\,meV, and a possible small gap $\Delta_S \approx 1.8$\,meV, and their temperature dependence. Additionally, a resonant coupling with a characteristic bosonic mode was observed. The boson energy at 4.2\,K, $\varepsilon_0 = 10.5-11.0$\,meV being less than the indirect gap ($\Delta_L < \varepsilon_0 < \Delta_L +\Delta_S$).

\end{abstract}


\maketitle

\section{Introduction}

Despite extensive investigations of the oxypnictide superconductors $Re$FeAsO ($Re$ is rare-earth metal) of the 1111 family since the discovery of iron-based superconductivity in LaFeAsO$_{1-x}$F$_x$ \cite{Kamihara}, many of its properties seem still ambiguous \cite{Si,Hirschfeld,HosonoRev}. Such problem arises from a lack of large enough single crystals, which makes the 1111 family hardly suitable for many techniques including angle-resolved photoemission spectroscopy (ARPES).

The layered crystal structure of the 1111 oxypnictides consists of quasi-two-dimensional superconducting Fe-As blocks separated by $Re$O spacers alternating along the $c$-direction. Being an antiferromagnetic metal at room temperature, the parent undoped compound $Re$FeAsO undergoes a structural and magnetic transition at $T^{\ast}$, thus turning to a spin-density wave (SDW) state below $T^{\ast}$. Being suppressed under electron or hole doping, SDW state gives a way to superconductivity. Unlike long known fluorine-substituted 1111, hydrogen-substituted $Re$FeAsO$_{1-x}$H$_x$ demonstrates double-dome superconducting state in the phase diagram \cite{Kobayashi,Fujiwara}.

Below $T_c$, two superconducting condensates are developed with the large gap $\Delta_L$ and the small gap $\Delta_S$ order parameters. Preliminary band-structure calculations \cite{Singh} showed several bands formed by Fe $3d$ orbitals crossing the Fermi level, with a formation of well-nested hole barrels near the $\Gamma$ point and electron barrels near M point of the 2-Fe Brillouin zone. For the 1111 family, due to a scarcity of momentum-sensitive probes of the superconducting order parameter, the gap distribution across the Fermi surface is still ambiguous, which differs dramatically from the situation with the 122 family.

Initial suggestion that the large gap $\Delta_L$ developed in the hole bands and the small gap $\Delta_S$ - in electron bands \cite{Mazin2008} was soon refuted for the majority of Fe-based superconductors. For now, $\Delta_S$ opening at the outer $\Gamma$ barrel, whereas a ``strong'' condensate with $\Delta_L$ developing in all other bands crossing $E_F$ is considered to be uniform for pnictides. As for oxypnictides, although such convention seems partly consistent with a few available ARPES data \cite{Charnukha1,Charnukha2}, further studies are obviously required. Additionally, a strong renormalization of the calculated band structure in SmFe$_{0.92}$Co$_{0.08}$AsO and NdFeAsO$_{0.6}$F$_{0.4}$ with critical temperatures $T_c = 18$\,K and 38\,K, respectively, was revealed \cite{Charnukha1,Charnukha2}, thus contradicting with general expectations \cite{Mazin2008}. This led to the band-edge singularities turned to a close proximity of the $E_F$ at $\Gamma$ and M points of the momentum space. Such nontrivial band picture, obviously unstable with respect to a fine tuning of the Fermi level, may cause featured densities of states (DOS) and carrier concentrations, in the bands where $\Delta_{L,S}$ are developed in the superconducting state.

In order to describe multiple band superconductivity in iron-based superconductors, several models were suggested: $s^{++}$-model of coupling through orbital fluctuations enhanced by phonons \cite{Nakaoka,Onari2017}, $s^{\pm}$-model of spin-fluctuation-mediated repulsion \cite{Hirschfeld,Mazin2008,Korshunov2011}, a shape-resonance model \cite{Bianconi}, and orbital-selective pairing \cite{Yu,Kreisel}. A spin resonance peak at the nesting vector was observed in neutron scattering probes \cite{Shamoto}. According to theory, the energy $\hbar\omega$ of spin exciton should fulfill the resonance condition $\hbar\omega < (\Delta_L + \Delta_S)$ or $\hbar\omega < 2\Delta_L$ \cite{Korshunov2016,Korshunov2018}.

A characteristic feature of heavily hydrogen-substituted 1111 is a sizable increase in $c$ lattice parameter which takes place in $x \rightarrow 0.5$ region of the phase diagram \cite{Hiraishi}. Such isostructural transition unaccompanied with AFM phase \cite{Saha} relates with non-nematic orbital fluctuations, which expected to gain $T_c$ within the second superconducting dome \cite{KontaniH}. From this point of view, the structure of the superconducting order parameter in H-substituted oxypnictides may perform novel and extraordinary features unlike other members of the 1111 family.

Here we present a direct probe of the superconducting order parameter in polycrystalline samples of hydrogen-substituted NdFeAsO$_{0.6}$H$_{0.36}$ (hereafter Nd-1111H) by using intrinsic multiple Andreev reflection effect (IMARE) spectroscopy. We determined the magnitudes of the two distinct superconducting order parameters and their temperature dependence, estimated intra- to interband coupling strength imbalance, and eigen parameters of both superconducting condensates (to be realized excluding interband coupling). A resonant coupling with a characteristic bosonic mode was observed, with the energy more than $\Delta_L$ and less than indirect gap $(\Delta_L + \Delta_S)$ at $T \rightarrow 0$ which satisfies the theoretical condition \cite{Korshunov2016,Korshunov2018}.

\section{Experimental details}

The polycrystalline sample with the nominal composition NdFeAsO$_{0.6}$H$_{0.36}$ was prepared in a cubic anvil high-pressure cell from the stoichiometric mixture of NdAs, FeAs, FeO, Fe, and Nd(OH)$_3$ powders. A pressure of 3~GPa was applied at room temperature. By keeping the pressure constant, the temperature was increased up to a maximum value of 1450~$^{\circ}$C, maintained for 14~h, followed by cooling to room temperature in 3~h. Overall details of the experimental setup can be found in our previous publications \cite{Zhigadlo2016,Zhigadlo2017}. X-ray measurements revealed the single-phase nature of the sample as well as the absence of a suitable amount of impurities. The occurrence of bulk superconductivity at critical temperature $T_c = 48$\,K was confirmed by the magnetic measurements.

In order to make superconductor - normal metal - superconductor (SnS) junctions for Andreev spectroscopy experiment, we used a break-junction technique \cite{Moreland,BJ}. The sample prepared as a thin rectangular plate with dimensions about $3 \times 1.5 \times 0.1$\,mm$^3$ was attached to a springy sample holder by four-contact pads made of In-Ga paste at room temperature. After cooling down to $T = 4.2$\,K, the sample holder was gently curved, thus cracking the bulk sample, with a formation of two cryogenic clefts separated with a weak link, a kind of ScS contact (where $c$ is a constriction). The resulting constriction turns far from current and potential contacts, which prevents junction overheating and provides true four-point probe. A layered sample splits along the $ab$-planes where steps and terraces naturally appear; the height of the step is a multiple of the $c$ unit cell parameter, whereas the terrace size appears about $10-100$\,nm. Typically this is the case for polycrystalline sample of layered compound as well. With regard to the 1111 family, highly expected is a number of cracked crystal grains with steps and terraces on its surface as shown by us earlier \cite{EPL,SmPRB,BJ}.

Under fine tuning the curvature of the sample holder, the two cryogenic clefts slide apart touching onto various terraces; they remain tightly connected during sliding that prevents impurity penetration into the crack and protects the purity of cryogenic clefts. Such tuning enables to sweep the constriction area in order to realize a desired ballistic regime (the contact dimension $d$ is less than the carrier mean free path $l$). In the majority of Fe-based superconductors we studied, the constriction is electrically equivalent to thin layer of normal metal of high transparency (about $95 \textendash 98 \%$),
thus providing an observation of multiple Andreev reflection effect (MARE). As a result, the obtained current-voltage characteristics (CVC) and the dI(V)/dV spectra are typical for the clean classical SnS-Andreev junction \cite{OTBK,Arnold,Averin,Kummel,Gunsenheimer}.

At temperatures below $T_c$, Andreev transport causes a pronounced excess current which drastically rises at low bias voltages (foot), and a series of dynamic conductance dips called subharmonic gap structure (SGS). At certain temperature, the position of SGS dips directly relates with the gap magnitude \cite{Kummel,Gunsenheimer}:

\begin{equation}
eV_n(T) = \frac{2\Delta(T)}{n},
\end{equation}
where $n=1,~2,\dots$ is natural subharmonic order. Unlike probing asymmetric NS and NIS junctions (I is insulator), no fitting of dI(V)/dV is need in case of SnS contact till $T_c$, which facilitates a precise measurement of temperature dependence of the gap. In principle, the first Andreev minimum could be slightly shifted towards lower bias, $V_{n=1} \lesssim (2\Delta/e)$ \cite{OTBK,Arnold,Averin,Kummel,Gunsenheimer}. If such happens, the gap value may be determined using the positions of the higher order SGS dips with $n \geqslant 2$. In case of two-gap superconductor, two SGS's are expected in the dynamic conductance spectrum.

For the junctions obtained in layered superconductors with a valuable anisotropy of electrical properties, the ballistics should be kept along both, $ab$ and $c$-directions. Here, due to the current flowing along the $c$-direction, the ballistic conditions are $l^{\rm inel}_c > d_c$, and $l^{\rm el}_{ab} > d_{ab}$, where $l^{\rm}$ and $l^{\rm inel}$ are the elastic and inelastic mean free path, $d_{ab}^2$ is the constriction area to be estimate using Sharvin formula \cite{Sharvin}. The number $n^{\ast}$ of observed Andreev dips corresponds with the beginning of the foot at the bias voltage $eV = 2\Delta/n^{\ast}$, and indirectly determines the out-of-plane inelastic mean free path $l^{\rm inel}_c$ to $d_c$ ratio: $n^{\ast} \approx 2l^{\rm inel}/d$~~~\cite{Kummel,Gunsenheimer} for the case of fully transparent contact.

\begin{figure}
\includegraphics[width=21pc]{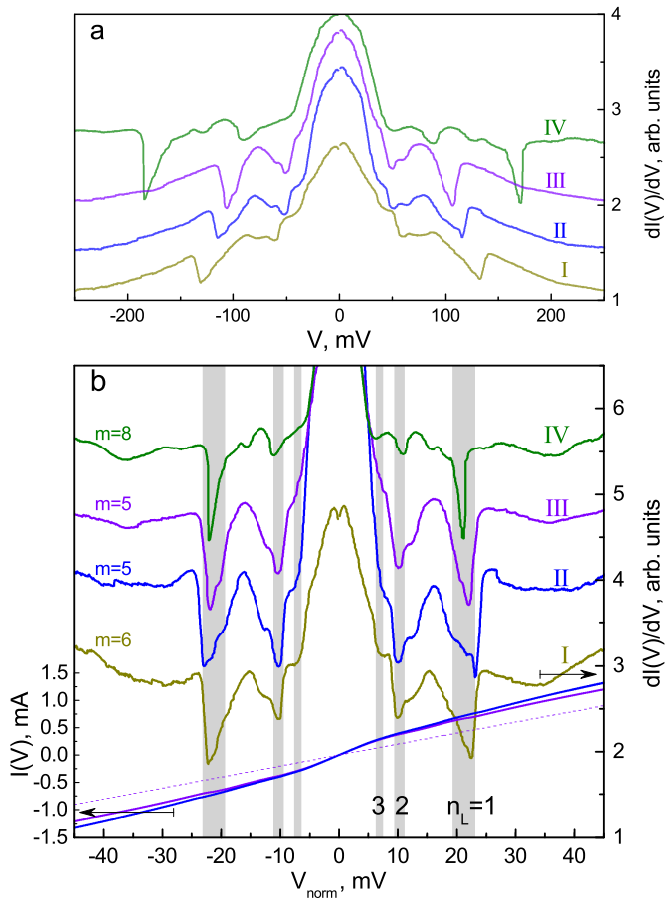}
\caption{(a) Raw dynamic conductance spectra measured at $T = 4.2$\,K for SnS Andreev arrays with various number of series junctions $m$. The curves were shifted vertically for clarity. (b) The same dI(V)/dV spectra with suppressed monotonic background; the bias voltage axis was normalized with corresponding $m$ ($V_{\rm norm} \equiv V_{\rm array}/m_i$). Also presented are current-voltage characteristics (left axis) for $m=5$ junction arrays. Dashed line shows simulated ohmic I(V) at $T_c$. Gray areas and $n_L=1,~2,~3$ labels indicate the position of subharmonic gap structure dips for the large gap $\Delta_L \approx 10.4$\,meV.}
\label{adj}
\end{figure}

Beside the single ScS contacts, Andreev arrays with ScSc-$\dots$-S structure can be also formed in the break-junction experiment with layered sample \cite{EPL,BJ,SmPRB}. Composed of $m$ ScS junctions, such array peers a natural stack of equivalent resistors (with parallel normal and Andreev channels). Hence, in the dI(V)/dV of the array, an intrinsic multiple Andreev reflection effect (IMARE) occurs, with the position of Andreev features scaled by a factor of $m$ as compared with that of single SnS junction:

\begin{equation}
eV_n(T) = m \times \frac{2\Delta(T)}{n}.
\end{equation}
IMARE is similar to intrinsic Josephson effect in SISI-$\dots$-S array observed in high-temperature cuprates and other layered superconductors \cite{PonomarevIMARE,PonomarevIJE,BJ,Nakamura}. Gently readjusting of the contact point, one can probe several tens of Andreev arrays with various diameter and number of junctions in one and the same sample and during the same cooldown. To the best of our knowledge, thus provided data statistics permits to check the data correctness in terms of reproducibility. For the formed array, the number $m$ is natural but accidental, so it can be determined by comparing dI(V)/dV curves for various arrays: after scaling the bias voltage axis by $m$, the dynamic conductance spectrum turns to that of a single junction. In Figs.~1-3,~5, each CVC and corresponding dynamic conductance spectrum is normalized using the determined $m$, thus corresponding to a single SnS contact. Hereafter, $V_{\rm norm}$ means $V/m$, whereas the current axis is kept unnormalized. The method of extracting $m$ numbers is detailed in the Appendix.

As method shows \cite{EPL,BJ}, Andreev dips in the dynamic conductance spectrum of array are more sharp and intensive than those for single SnS junction; the larger $m$, the sharper dI(V)/dV features. This firm experimental fact favors a natural origin of such arrays (as a part of layered structure) rather than a chain of independent nonequivalent grain-grain contacts \cite{remark_1}.

During (I)MARE, electron could lose or gain its energy by coupling with some bosonic mode. At low temperatures, boson emitting seems more likely, whereas the energy of the bosonic mode $\varepsilon_0$ is up to $2\Delta$. A resonant interaction with a characteristic bosonic mode with peculiar energy $\varepsilon_0$ causes a fine structure in the dI(V)/dV spectrum. Accompanying each Andreev dip, at higher bias, less-intensive satellite dip appears at position

\begin{equation}
eV_n = \frac{2\Delta+\varepsilon_0}{n},
\end{equation}
forming an additional subharmonic series. The resulting fine structure looks similar as compared with the case of microwave irradiated SnS junction observed firstly in YBaCuO \cite{Zimmermann}.

Summarizing the advantages of IMARE spectroscopy of break-junctions and natural arrays, this technique provides a precise and high-resolution probe of the bulk superconducting order parameter, its temperature dependence and fine structure. In our studies, the dynamic conductance spectra were measured directly by a standard modulation technique \cite{LOFA,BJ}. We used a source of direct current mixed with a small-amplitude $ac$ with frequency about 1\,kHz from the external oscillator. The results obtained with this setup are insensitive to the presence of parallel ohmic conduction paths; if any path is present, the dynamic conductance curve shifts along the vertical axis only, while the bias stay unchanged.

\section{Results and discussion}

\subsection{Subharmonic gap structures}
\begin{figure*}
\includegraphics[width=\textwidth]{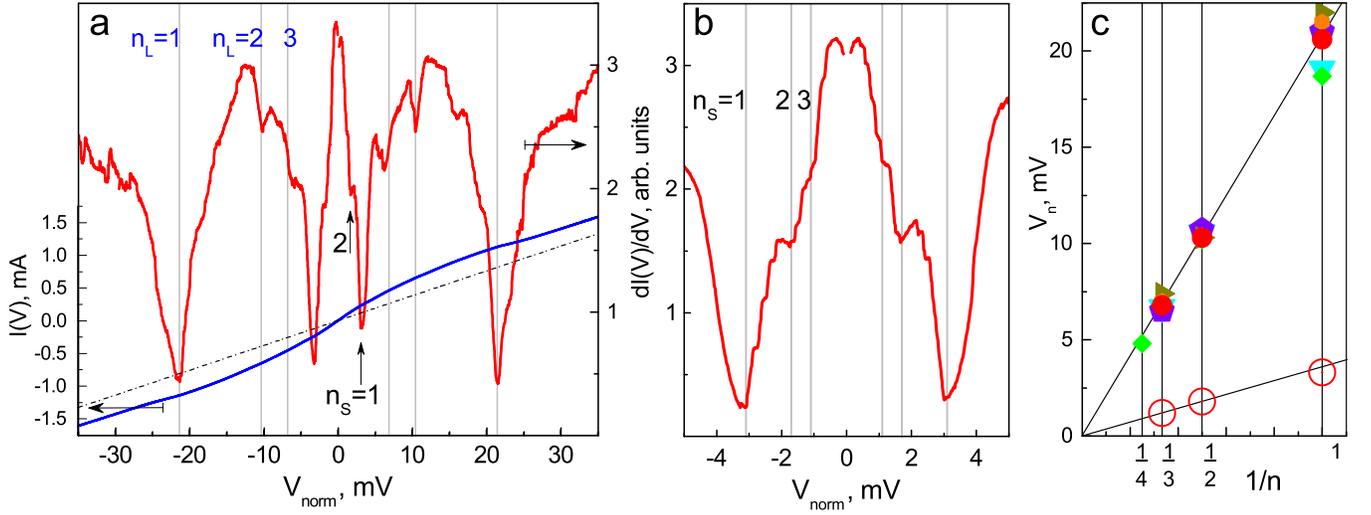}
\caption{a) Dynamic conductance (right axis) at $T=4.2$\,K of Andreev array ($m=12$ junctions) showing two subharmonic gap structures: for the large gap $\Delta_L \approx 10.5$\,meV (vertical gray lines and blue $n_L=1$,~2,~3 labels), and for the small gap $\Delta_S \approx 1.8$\,meV (arrows and $n_S$ labels). $V_{\rm norm} \equiv V_{\rm array}/12$. Monotonic background is suppressed for clarity. Current-voltage characteristic (left axis) at $T = 4.2$\,K and its simulation at $T_c$ (dash-dot line) are shown for comparison. b) The low-bias fragment of the dI(V)/dV spectrum shown in (a) which details the Andreev structure of the small gap (vertical gray lines, $n_S=1$,~2,~3 labels). Monotonic background is suppressed separately for positive and negative bias. c) The SGS positions versus their inverse number $1/n$ for the large gap (solid symbols) and the small gap (open symbols) in dI(V)/dV spectra of various Andreev arrays shown in Figs.~1-3,~5. The data in (a,b) panels are illustrated with red circles. Solid lines are guidelines.}
\label{sgap}
\end{figure*}

Figure~1a shows typical raw dynamic conductance spectra of the break-junctions formed in Nd-1111H samples at $T=4.2$\,K. The spectra demonstrate an excess conductance which rises toward low bias voltages, and a series of Andreev dips. The position of the dips is although irreproducible, since the spectra correspond to the arrays with various number of junctions, thus providing $\Delta \cdot m$ energy value. In order to reduce each spectrum to that of a single junction, the bias voltages were divided by $m = 6,~5,~5,~8$, correspondingly (from the bottom). How these $m$ were chosen, see the Appendix.

The normalized CVCs (for II and III spectra), as well as the spectra from Fig.~1a with suppressed monotonic background, are shown in Fig.~1b by similar colours. Normal-state CVC for the III spectrum at $T=T_c$ simulated with dashed line, determines the normal resistance $R_N \approx 50~{\rm \Omega}$ per junction. As compared with ohmic dependence, the CVC measured at 4.2\,K demonstrates a pronounced excess current. In order to check the ballisticity of the constriction in the $ab$-plane, we take the following parameters. For a single crystal grain, the normal state in-plane resistivity $\rho^{ab}(T_c) = 0.13-0.15~{\rm m\Omega \cdot cm}$ is similar to that of other oxypnictides synthesized in the same way \cite{Zhigadlo2008,Zhigadlo2010,Zhigadlo2012}, whereas the in-plane coherence length at $T \rightarrow 0$ is $\xi^{ab}(0) \approx 2.1$\,nm. Using the in-plane Ginzburg-Landau penetration depth for sister Sm-1111 single crystal with similar $\rho(T_c)$ from \cite{Weyeneth} $\lambda_{GL}^{ab}(0) \approx 200$\,nm, we determine the clean-limit value $\lambda_L^{ab} \approx 195$\,nm. Taking the average Fermi velocity \cite{Singh} $v_F \approx 1.4 \times 10^8$\,cm/s, we get the $ab$-plane product of the bulk resistivity and the elastic carrier mean free path $\rho l^{el}=\mu_0\lambda_L^2 v_F \approx (6.6-6.7) \times 10^{-7}~{\rm m\Omega \cdot cm^2}$, and therefore estimate $l^{el} \approx 44-52$\,nm for single crystal of Nd-1111H. Finally, the diameter of the constriction \cite{Sharvin} $2a = 2\sqrt{4\rho l^{el}/3\pi R_N} \approx 24$\,nm is nearly 2 times less than estimated elastic mean free path, thus proving the junction to be ballistic, with 2-3 Andreev subharmonics expected in the dI(V)/dV spectrum. Generally speaking, a measure of ballisticity is not elastic but inelastic mean free path to $2a$ ratio, to appear an order of magnitude higher than estimated $l^{el}/2a$.

The beginning of the drastic rise of dynamic conductance at low bias (foot) roughly matches the position of $n_L = 3$ subharmonic of the large gap, therefore, $n^{\ast} \approx 2 l^{\rm inel}_c/d_c \approx 3$ (both characteristic lengths are taken along the $c$-direction) \cite{Gunsenheimer}. As a result, the ballistic along the $c$-direction is also satisfied: $l^{\rm inel}_c/d_c \approx 1.5$ for the case of fully transparent constriction. In case of high but finite transmission probability ($0.1-0.2$) this ratio tends to 2. The above estimates in the in-plane and out-of-plane directions signify $2-3$ SGS dips are expected in the dynamic conductance spectra. Additionally, a pronounced excess current indicates a high-transparency Andreev mode.

The arrays I, II and III were formed in one and the same sample sequently. Under fine tuning of the holder curvature, the initial contact point onto a 6-junction stack (curve I) jumped to a neighbour terrace onto a 5-junction stack (curve II), then swept (curve III) thereby changing the area and resistance of the junction. However, despite such metamorphosis, the normalized dynamic conductance spectra look quite similar. Pronounced dips at $eV \approx \pm 21,~\pm10.4$\,meV and shoulders at $\pm 6.9$\,meV being $n_L=1,~2,~3$ subharmonics comprise SGS of the large gap $\Delta_L \approx 10.4$\,meV. For comparison, the dI(V)/dV spectrum for 8-junction array formed in the next sample from the same batch (curve IV) is shown in Fig.~1 as well. Remarkably, the position of the gap subharmonics is insensitive to the topology of the contact point onto the cryogenic surface, and to be reproducible from one sample to another. Therefore, the extracted $\Delta_L$ is a bulk order parameter independent on the contact dimension and almost unaffected with a surface proximity.

The CVC and the dynamic conductance spectrum which demonstrates two distinct SGS are shown in Fig.~2a. $n_L$ labels point to Andreev dips of the large gap $\Delta_L \approx 10.5$\,meV. At lower bias, another set of features presents (up arrows), beginning with pronounced dip at $eV_{n_S=1} \approx \pm 3.3$\,meV being much more intensive than the third-order $n_L=3$ feature of the large gap. The position of this feature does not match that of the fourth-order subharmonic of the $\Delta_L$ expected at $\pm 5.5$\,meV. The fragment which details the low-bias region of the dI(V)/dV curve is shown in the inset. After additional suppressing of monotonic background, minima at $|eV_{n_S=1}| \approx 3.3$\,meV, $|eV_{n_S=2}| \approx 1.8$\,meV and shoulders at $\pm 1.2$\,meV become clear, all together could be interpreted as SGS of the small gap energy parameter $\Delta_S \approx 1.8$\,meV.

The above interpretation seems for us the most reliable. Yet to be discussed are the other possible origins of such low-bias SGS (see open circles in Fig.~2c).

1) Supposing the proximity effect origin, with a bulk order parameter inducing on the cryogenic cleft a formation of Cooper pairs with bond energy $\Delta_{\rm surf} < \Delta_{\rm bulk}$, one have to use the raw bias (unnormalized by $m$) in order to determine this surface order parameter. For the case, at 4.2\,K the raw position of the feature pointed by up arrow in Fig.~3a is $\approx 39.6$\,mV which corresponds to $\Delta_{\rm surf} \approx 20$\,meV. Obviously, being much larger than $\Delta_L$, such order parameter cannot be induced, thus making the considered case improbable.

2) As could be numerically shown in the framework of K\"{u}mmel \textit{et al.} theory \cite{Kummel}, if Andreev bound states (ABS) appear in the constriction, with energies $\varepsilon_{\rm ABS} < \Delta_{\rm bulk}$, the dynamic conductance spectrum shows additional subharmonic structure at positions $(\Delta_{\rm bulk}+ \varepsilon_{\rm ABS})/en$. Let us find a way to reproduce the observed low-bias structure. If $d_c \gg 10 \xi_c(0)$, then relative to $E_F$, for the first Andreev level $\varepsilon_{\rm ABS} \rightarrow 0$. Thus, the additional structure caused by the in-gap ABS starts from $\approx \Delta_{\rm bulk}/e$ bias which nearly coincides with the position of $n_L = 2$ dip being almost three times greater than observed. The second minimum related to the ABS structure expected at $\Delta_{\rm bulk}/2e$ is absent in the presented spectra. Therefore, this low-bias feature set cannot be caused by ABS.

3) Consider superconducting gap anisotropy in the $k$-space or an appearance of parasitic junction in parallel. Both suggestions would cause an additional SGS which starts from unpredictable bias $E^{\ast}$ but evolves with temperature similarly to $\Delta_L(T)$. Nonetheless, the two observed structures behave differently (see $V_{n_S}(T)$ normalized by $V_{n_S}(0)/V_{n_L}(0)$ dependence shown by crosses in Fig.~3b), making these cases unrealistic.

4) In case of asymmetric junction SnS$^{\ast}$, where S$^{\ast}$ is a single-gap superconductor with slightly reduced order parameter (for example $\Delta^{\ast} \approx 5/6 \Delta_L$), the dynamic conductance spectrum would show two subharmonic structures at $(\Delta_L \pm \Delta^{\ast})/en$. Therefore, the main $n=1$ features are to appear at $11/6 \cdot \Delta_L \approx 2\Delta_L$ and $\Delta_L/6$. The position of the $(\Delta_L - \Delta^{\ast})/en$ structure then suits the observed one. However, such tunneling transition is forbidden at $T \rightarrow 0$, whereas its probability increases with temperature. Would it be the case, the dI(V)/dV features to be intensify toward $T_c$. The experimental spectra in Fig.~3a (see the inset for details) demonstrate the opposite tendency. Yet in the spectrum measured at 17\,K, the minima almost vanish, thus to be distinguished after background suppressing only.

\begin{figure}
\includegraphics[width=20pc]{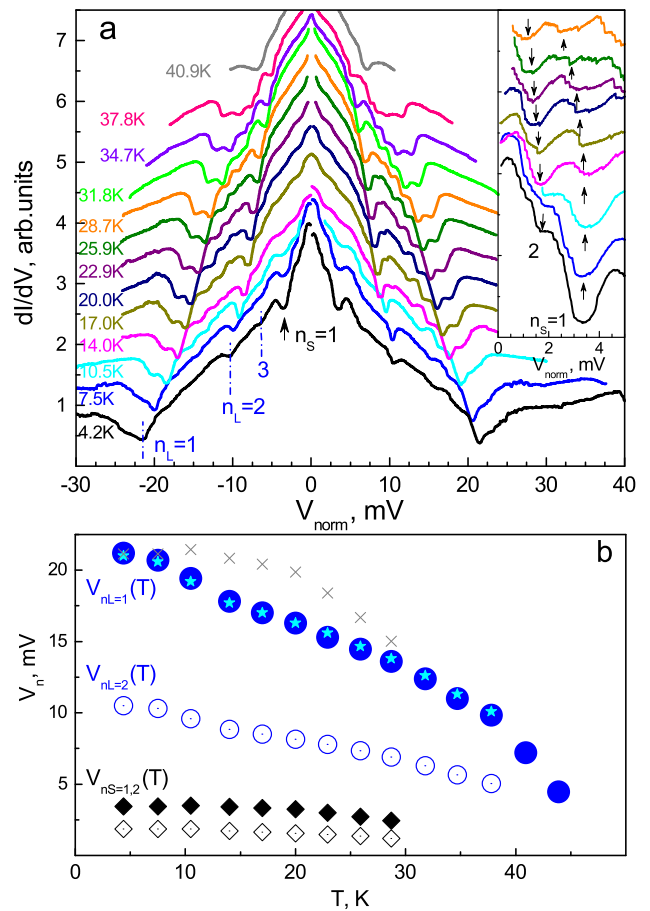}
\caption{a) Evolution of the dynamic conductance spectrum with temperature for SnS Andreev array from Fig.~2a. Dash-dot bars and $n_L=1,~2,~3$ labels indicate the subharmonic gap structure dips for the large gap $\Delta_L \approx 10.5$\,meV. The inset shows the low-bias fragments (with suppressed monotonic background) of the spectra shown in (a) which detail the first ($n_S=1$, up arrows) and the second ($n_S=2$, down arrows) features for the expected small gap $\Delta_S \approx 1.8$\,meV. dI(V)/dV curves are shifted vertically for clarity. b) Temperature dependence of the first (solid symbols) and the second (open symbols) Andreev features $V_n(T)$ for the large gap (circles), and for the small gap (rhombs). Stars and crosses show how $2\cdot V_{n_L=2}(T)$ and $V_{n_S=1}(T)$ dependences, respectively, correspond with $V_{n_L=1}(T)$ for the large gap SGS.}
\label{temp}
\end{figure}

The position of the gap features in the dynamic conductance curves of various Andreev arrays versus their inverse number is summarized in Fig.~2b. For the large gap (solid symbols), the resulting linear dependence tending to the origin agrees well with Eq.~1. Therefore, the large gap order parameter of the average magnitude $\Delta_L \approx 10.4$\,meV is observed reproducibly in the Andreev spectra of Nd-1111H. The less-sloped line fits the positions of the second set of SGS, which determines the characteristic energy $\approx 1.8$\,meV (open symbols). By analogy with other 1111 compounds \cite{EPL,UFN2014,SmPRB}, we suppose this energy corresponds to the small gap $\Delta_S$.

\subsection{Gap temperature dependence}

Temperature evolution of the dynamic conductance of Andreev array from Fig.~2a (original background is preserved) is shown in Fig.~3a. In the spectrum measured at 4.2\,K, the SGS dips of the large gap $\Delta_L \approx 10.5$\,meV are marked with dashed bars. In order to detail the SGS of the small gap $\Delta_S \approx 1.8$\,meV, in the inset we show the low-bias fragments of the spectra with suppressed monotonic background. Here, up arrows point to the first $n_S=1$ dip position at various temperatures, down arrows -- to the second one. As temperature increases, all gap features move towards zero bias. For the large gap, the position of the first (circles) and the second (open circles) dips versus temperature is presented at Fig.~3b, and directly associated with the $\Delta_L(T)$ temperature dependence (Fig.~4). The large gap trend looks non-typical as compared with other 1111 oxypnictides we studied \cite{UFN2014,SmPRB}. Nonetheless, when doubling the position of the second feature (originally located at $eV_2(T)=\Delta(T)$), it turns to that of the first one at all temperatures till $T_c$ (stars in Fig.~3b). On the one hand, such correspondence indicates that the observed features belong to one and the same SGS. On the other hand, it proves their noticeably curved temperature dependence originates from intrinsic superconducting phenomena natural to Nd-1111H rather than any undesired force (for example, local excess of the critical current density would lead to $V_{n_L=1}$ ``overheating'', i.e. $V_{n_L=1} < 2 V_{n_L=2}$).

\begin{figure}
\includegraphics[width=20pc]{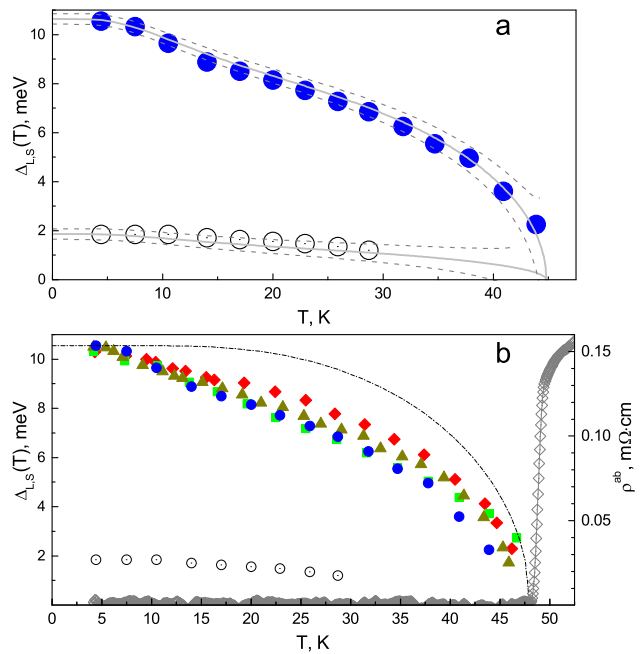}
\caption{Temperature dependence of the superconducting gaps obtained a) using data in previous figure, and b) compared with other $\Delta_L(T)$ measured with the samples from the same batch. $\Delta_L(T)$ dependences are shown by solid symbols, $\Delta_S(T)$ -- by open circles. Theoretical fits using two-band RBCS model are shown with solid lines (a), dash-dot line corresponds to a single-band BCS-like behaviour (b). Gray dashed lines frame the confident intervals for the large and the small gap values. Resistive superconducting transition of the dense multicrystalline material is presented by gray open rhombs (right scale).}
\label{gapt}
\end{figure}

Using the $V_n(T)$ dependences (Fig.~3b), we obtain the temperature dependence of the large (solid circles) and the small gap (open circles) as an average between the SGS positions $\langle\Delta(T)\rangle = [V_{n=1}(T)+2V_{n=2}(T)]/4$ shown in Fig.~4a. The large gap dependence passes well below the single-band BCS-like curve: $\Delta_L(T)$ drops down at $T \approx 12$\,K, then decreases gradually, turning to zero at local critical temperature (the temperature of the contact area transition to the normal state). Such unusual behaviour reproduces the $\Delta_L(T)$ dependences measured using data with other Nd-1111H samples from the same batch (triangles, rhombs and squares in Fig.~4b), hence being independent on the resistance of the constriction, or current density through the junction. Nonetheless, despite the large data statistics obtained for Nd-1111H (more than 100 SnS arrays), only once we managed to get two SGS's for $\Delta_L$ and $\Delta_S$ up to $T_c$ (shown in Fig.~3). In the majority of the obtained dI(V)/dV, the Andreev features of the small gap are undistinguishable (for example, see Fig.~1). However, one should not think the obtained $\Delta_S(T)$ dependence as an artifact: indirectly, our data favors the existence of the second order parameter, whereas an exact experimental reason seems causing the strongly smeared SGS of the small gap in the obtained spectra, as discussed below.

The presumed small gap decreases more regularly with temperature increase. Obviously, the different temperature trend indicates that the resolved energy parameters relate with two distinct superconducting condensates coexisting in Nd-1111H. The characteristic ratio for the large gap $2\Delta_L/k_BT_c = 5.0-5.4$ exceeds the BCS-limit 3.5, whereas for the small gap $2\Delta_S/k_BT_c \approx 0.9 \ll 3.5$, seemingly caused by interband interaction.

The obtained $\Delta_{L,S}(T)$ are fitted with a two-band BCS-like model based on Moskalenko and Suhl equations \cite{Mosk,Suhl} using a renormalized BCS integral. In order to obtain the theoretical $\Delta_{L,S}(T)$ curves, beside the experimental values of $T_c$, $\Delta_{L,S}(T)$, we used $\alpha = \lambda_{LS} /\lambda_{SL}$, the ratio between effective intraband and interband couplings $\beta \equiv \sqrt{\lambda_{LL} \lambda_{SS}/(\lambda_{LS} \lambda_{SL})}$ ($\lambda_{ij}$ are reduced coupling constants extracted directly from the fit), and eigen temperature renormalization coefficient for each band as variables. Note the normal DOS at the Fermi level imbalance $N_S(0)/N_L(0) \neq \alpha$ (in order to get $N_S/N_L$, one should use full coupling constants, see for example page 2 in \cite{KGM}). The resulting solid lines in Fig.~4a fit the case of a strong interband coupling comparable with the intraband one, with $\beta \approx 1.5$. Taking the cutoff energy $\hbar \omega_{\rm cut} = 40$\,meV and $\ln (E_F/\hbar \omega_{\rm cut}) = 2$, we obtain the set of renormalized coupling constants $\lambda_{LL} = 0.33$, $\lambda_{SS} = 0.14$, $\lambda_{LS} = 0.49$, $\lambda_{SL} = 0.041$, which indicates $\lambda_{LS}$ dominating over the other pairing channels. The notable curvature of $\Delta_L(T)$ seems caused by large $\alpha$ value. Additionally, the eigen superconductivity of the $\Delta_L$ bands (to be realized in case of $\lambda_{LS}=\lambda_{SL}=0$) is close to the weak-coupling limit with $[2\Delta_L/k_BT_c]^{\rm eigen} \approx 3.5$ and eigen critical temperature $T_L^{\rm eigen} \approx 26.5$\,K. In contrast, a moderate coupling develops in the $\Delta_S$ bands with $[2\Delta_S/k_BT_c]^{\rm eigen} \approx 4$ and eigen critical temperature $\approx 0.33$\,K.

\begin{figure}
\includegraphics[width=20pc]{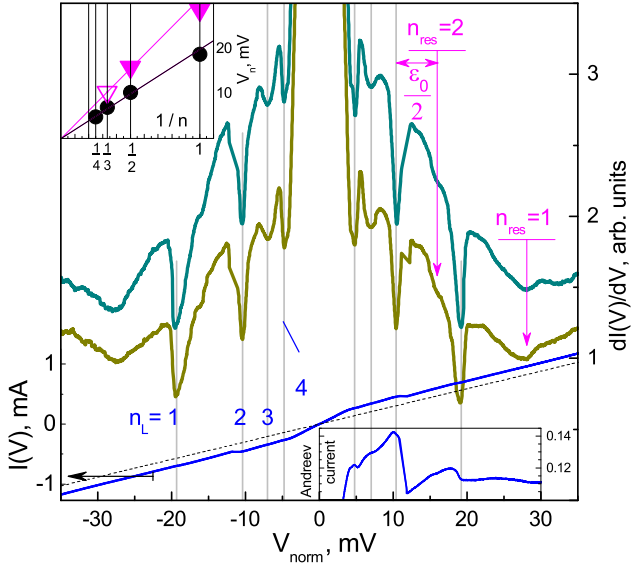}
\caption{Dynamic conductance spectra (right scale) measured at $T = 4.2$\,K for Andreev arrays ($m=14$ junctions). $V_{\rm norm} \equiv V_{\rm array}/14$. Gray lines and $n_L =1-4$ labels indicate the subharmonic gap structure dips for the large gap $\Delta_L \approx 10.3$\,meV. Vertical arrows point to the fine structure dips caused by resonant emission of bosons with energy $\varepsilon_0 = 10.5-11.0$\,meV. CVC at $T=4.2$\,K (blue line, left scale) and its simulation at $T_c$ (dashed line) are shown for comparison; the Andreev current component is presented in the lower inset. The upper inset shows the position of gap features (circles) and fine structure (triangles) versus their inverse number. Solid lines are guidelines.}
\end{figure}

As a rule, the cutoff energy $\hbar \omega_{\rm cut}$, generally ambiguous, is taken such as to fit the value of critical temperature. Here, $T_c$ is determined experimentally, therefore the chosen $\hbar \omega_{\rm cut}$ determines the values of the coupling constants, nevertheless without affecting the curvature of the theoretical $\Delta_{L,S}(T)$ fits. Under $\omega_{\rm cut}$ variation, unchanged remain also $[2\Delta_S/k_BT_c]^{\rm eigen}$, $\alpha$, and $\beta$: being the manually adjustable parameters, they are determined in terms of the best fit of the experimental data. Varying significantly the cutoff energy, $\hbar \omega_{\rm cut}$ even by $\pm 25\%$,
we get the $T_L^{\rm eigen}$ deviation $\pm 6\%$,
which remains almost stable. The only value notably modified under such $\hbar \omega_{\rm cut}$ change is the eigen critical temperature of the small gap: its variation is nonlinear and appears as high as $-30\%$ to $+40\%$,
thus making the absolute value of $T_S^{\rm eigen}$ ambiguous.

\subsection{Resonant coupling with a bosonic mode}

Beside the parent SGS, in the most qualitative dynamic conductance spectra, we resolved a fine structure caused by a resonant boson emission along with MAR process. Fig.~5 shows CVC and dI(V)/dV curves for two sequently formed Andreev arrays, with clear SGS for the large gap (gray vertical lines). The lower inset shows an excess Andreev current taken as I(V)-I$^{\rm Ohmic}(V)$. The intensive dips at bias voltages 19.2, 10.3, 6.9, and 5.1\,mV well satisfy Eq.~1 (accounting the first dip slightly shifted toward less bias). The dip positions versus $1/n$ (circles in the inset) follow a line passing through the origin and determine $\Delta_L \approx 10.3$\,meV.

Accounting the dip at 5.1\,mV is more pronounced than that at 6.9\,mV, it could be attributed to the first dip $n_S=1$ of the small gap $\Delta_S \approx 2.6$\,meV. In that case, the absent higher-order $\Delta_S$ subharmonics ($n_S = 2$ dip is expected at 2.6\,mV) could be considered as masked by a strong foot. Nonetheless, the supposed $\Delta_S$ value appears about 1.4 times larger than that extracted from Figs.~2,~3, thus such attribution ambiguous. Instead, we interpret the dip at 5.1\,mV as relating to the large gap SGS ($n_L=4$), whereas the beginning of the foot seems intensifying this dip as matching its position.

Accompanying the SGS for the $\Delta_L$, less intensive dips at $V_{\rm res} \approx 28.4,~15.8$\,mV appear in Fig.~5, which resembles a typical fine structure observed in precedent IMARE probes with Gd and Sm-based oxypnictides \cite{Gdboson,EPLboson}. The satellite position is independent on the constriction dimension or resistance, thus possibly originates from a resonant coupling with a characteristic bosonic mode. Earlier we showed \cite{Gdboson,EPLboson}, the fine structure cannot be caused by a electron-phonon interaction or Leggett mode.

The satellite position (vertical arrows) in dependence of inverse number is shown in the inset (triangles). The boson energy harmonics $\varepsilon_0/n$ are therefore the ``distances'' between each satellite and the parent SGS dip (Eq.~3). Accounting the first $\Delta_L$ dip shifts to bias lower than $2\Delta/e$, which entails possibly nonlinear shifting of the first bosonic resonance feature ($n_{\rm res}=1$ in the main panel of Fig.~5), it would be more correct to probe $V_{\rm res2}-V_{n_L=2} \equiv \varepsilon_0/2$ (double arrow in Fig.~5) in order to estimate the boson energy $\varepsilon_0 = 10.5-11.0$\,meV. Extrapolating the obtained $V_{\rm res}(1/n)$ dependence, we get the expected position $\approx 10$\,mV of the third satellite (open triangle in the inset), which fits the position of the second gap subharmonic $n_L=2$. Such overlapping seems the reason for $n_{\rm res}=3$ unresolved as distinct feature.



The estimated boson energy at $T \rightarrow 0$ is less than the indirect gap $\varepsilon_0 < \Delta_L(0) + \Delta_S(0)$, and, in accordance with theory \cite{Korshunov2016,Korshunov2018}, supports a spin-exciton nature of the observed boson. The energy which similarly fulfills the resonance condition was extracted by us earlier in other 1111 compounds within a wide range of critical temperatures \cite{Gdboson,EPLboson}. The measurements of the temperature dependence $\varepsilon_0(T)$ are necessary in order to clarify this phenomenon nature. Also, further studies are need to check the result reproducibility.

\subsection{Discussion}

Summarizing the experimental data, our IMARE studies of Nd-1111H unambiguously show a presence of the bulk superconducting order parameter $\Delta_L = 10.45 \pm 0.15$\,meV. Its characteristic ratio 5.0-5.4 joins the large statistics for oxypnictides with substitution sites in the spacer, collected by us earlier \cite{UFN2014,SmPRB}.

Some dI(V)/dV measured with Nd-1111H demonstrate a second set of Andreev features at lower bias, which could be referred to as SGS for the small gap $\Delta_S \approx 1.8$\,meV. Although the small gap features are poorly observed in the dI(V)/dV spectra, several arguments could be shown pro and contra the presence of the second superconducting condensate with $\Delta_S$ order parameter:

1. If the observed $\Delta_L$ were a single order parameter, its temperature dependence would be trivial. On the contrary, reproducibly observed curvature starting at $\approx 12$\,K (Fig.~4) cannot be simulated in any conventional single-gap model. In principle, the only way to reproduce the curved $\Delta(T)$ within a single-gap approach is as follows. Relying on the ARPES data \cite{Charnukha1,Charnukha2}, plausible seems an existence of the flat bands in the vicinity of $E_F$, with extremely high carrier effective mass and DOS. Such nontrivial band structure could be sensitive not only to doping, but also to the temperature. In the superconducting state, such possible band shifting would cause a DOS at $E_F$ temperature dependence $N_0(T)$. Even weak $N_0(T)$ alteration would result in relatively strong temperature dependence $\lambda(T)$, thus making it not a constant (always eliminated in conventional models) and providing unconventional behaviour of the superconducting gap even within a single-gap approach.

2. On the other hand, the obtained temperature dependences (Fig.~4) are natural for conventional two-gap model. In this framework, the reason of the curved $\Delta_L(T)$ is a cooperation of $N_S/N_L > 1$ and a moderate interband interaction in the momentum space with a ``driven'' superconducting condensate.

3. In general, it is the $\Delta_S$ band contribution to Andreev conductivity which manages the observability of the small gap SGS in dI(V)/dV spectrum. Evanescent band structure supposed near the Fermi level \cite{Charnukha1,Charnukha2} is generally followed with low carrier concentration and possibly high DOS. In this sense, poorly observable SGS of the small gap may result from a lower concentration or mean free path for carriers from those bands, as compared with the bands where $\Delta_L$ develops.

From the experimentalist point of view, only the reproducible observation of the small gap features seems a key to above mentioned issue.

The outlined ambiguity in the small gap of Nd-1111H resembles the situation with another optimally doped oxypnictides having similar critical temperatures, studied previously \cite{EPL,UFN2014,SmPRB}. We showed, subtle for optimally doped (Sm,Th)OFeAs, the $\Delta_S$ Andreev features become more pronounced for underdoped samples as critical temperature decreases; with it, the small gap magnitude scales with $T_c$ along with the large gap \cite{SmPRB}.


\section{Conclusions}

By using intrinsic multiple Andreev reflections effect (IMARE) spectroscopy, we probed the structure of the superconducting order parameter in polycrystalline samples of hydrogen-substituted NdFeAsO$_{0.6}$H$_{0.36}$ compound with a critical temperature $T_c = 45-48$\,K. At 4.2\,K, we determined the two bulk superconducting gaps: the large gap $\Delta_L \approx 10.4$\,meV, and a possible small gap $\Delta_S \approx 1.8$\,meV. Supposing a two-gap scenario, the temperature dependence of the gaps could be fitted within a renormalized BCS two-band approach. We estimate the large gap condensate is in a weak coupling limit, while a moderate pairing is developed in the bands with the small gap. The interband constant $\lambda_{LS} \approx 0.5$ dominates over the other pairing channels. Additionally, we revealed a resonant coupling of the Andreev current with a characteristic bosonic mode with energy $\varepsilon_0 = 10.5-11.0\,{\rm meV} < \Delta_L + \Delta_S$ at $T = 4.2$\,K.

\begin{acknowledgments}
We acknowledge the support from RFBR (project no. 17-02-00805). The research has been partly done using the research equipment of the Shared facility center at Lebedev Physical Institute RAS.
\end{acknowledgments}

\appendix*
\section{IMARE spectroscopy details}
\begin{figure*}
\includegraphics[width=\textwidth]{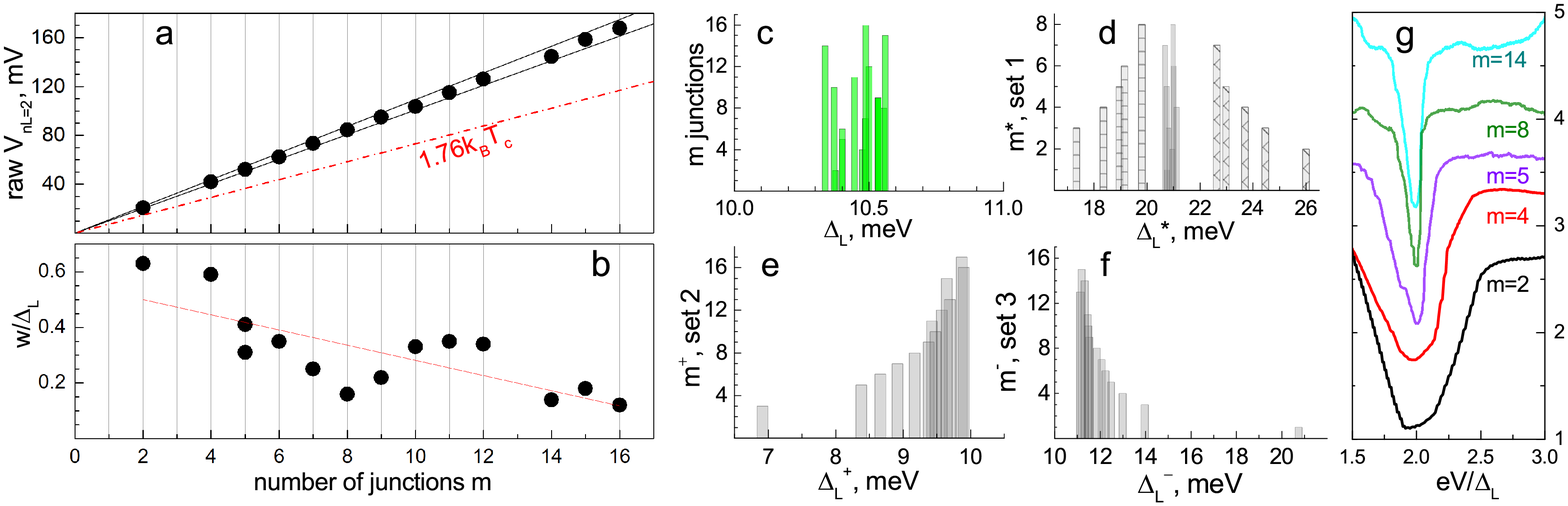}
\caption{(a) The raw positions of the second Andreev subharmonic $V_{n_L=2} = \Delta_L/e$ in the obtained dI(V)/dV spectra versus the congruent $m$ numbers of junctions in the stack. The dash-dot line designates the BCS-limit $1.76 k_BT_c$ below which the experimental points not to be. (b) The half-width of the main $n_L = 1$ Andreev dips normalized to $\Delta_L$ in dependence on the number of junctions. Solid line is a guideline. The extracted $\Delta_L = 10.45 \pm 0.15$\,meV values are shown in (c). For comparison, the large gap values to be obtained when the raw $V_{n_L=2}$ are normalized with another sets of $m^{\ast}_{\rm even}=m/2$ (gray bars), $m^{\ast}_{\rm odd} = m/2 \pm 0.5$ (horizontally and diagonally dashed bars, respectively), $m^+ = m+1$, and $m^- = m-1$, are shown in (e-f) panels. The bar height corresponds to the chosen $m$. (f) The main Andreev dips ($n_L=1$) observed at $2\Delta_L/e$ bias in the representative spectra of Andreev arrays with various $m$. The dI(V)/dV curves are shifted vertically for clarity, monotonic background was suppressed.}
\label{hist}
\end{figure*}

When cracking a sample with a layered crystal structure, array contacts are naturally develop on the steps and terraces of the cryogenic cleft. Although spectroscopy of such stacks seems rather advantageous, as compared with single junction study, one more intermediate purpose raises, namely to determine the number of junctions in each formed array. Primarily, in order to solve this problem, a large data statistics is essential. Above it was mentioned, the raw position of the Andreev feature becomes scaled by a factor of natural but accidental $m$, hence, the raw dI(V)/dV spectrum provides $\Delta \cdot m$ energy value. The details of the $\Delta$ value extraction from the raw experimental data are presented in Fig.~6(a-f).

One possible way is to arrange in ascending order the positions of the second Andreev feature $V_{n_L=2} = m\Delta_L/e$ in the obtained raw dI(V)/dV spectra, then to assign them natural numbers in order to get a straight line crossing the origin (Fig.~6a). Of course, the obtained points should be held above the BCS-limit line $1.76k_BT_c$. For such $V_n(m)$ estimate, the position of the second ($n=2$) Andreev subharmonic seems more suitable since exactly corresponding $\Delta$ (by contrast, the position of the first subharmonic could be a bit lower $2\Delta/e$ \cite{Kummel,Gunsenheimer}). Fig.~6a shows the dependence of raw $n_L=2$ positions versus the selected $m$ numbers well-fitted with a line.

In (c) panel, the extracted values of $\Delta_L$ are shown (i.e. the raw position shown in (a) divided by the selected $m$). Each bar represents a certain array, with the bar height corresponding to the number of junctions in the array. Remarkably, the selected set of $m$ provides the gap value uncertainty less than $2\%$, thus proving the selected $m$ set is correct.
Noteworthily, there is no correlation between the gap value and the corresponding $m$, which indicates to a bulk nature of the extracted order parameter.

For comparison, the gap values to be obtained when using another sets of $m$ are shown in Fig.~6(d-f). Panel (d) depicts the gap distribution $\Delta_L^*$ to be obtained if assuming the $m=2$ junctions array to be a single junction with $m^{\ast}_{\rm even}=m/2$ (gray bars), $m^{\ast}_{\rm odd} = m/2 \pm 0.5$ (horizontally and diagonally dashed bars, respectively). Under such normalizing, the contacts with even $m$ provide reproducible and non-scattered but double $\Delta_L$ value (gray bars), whereas the odd $m$ numbers are to be rounded, thus providing ambiguous gap value (dashed bars). Panels (e,f) expose the distributions following $m^+ = m+1$ and $m^- = m-1$, respectively. Clearly, (e-f) cases provide strongly scattered gap values, with obvious tendency $(\Delta_L^+,~\Delta^-) \rightarrow \Delta_L$ with $m^{+,-}$ increase (d,e), thus supposing badly wrong $\Delta_L^{+,-}$ correlation with the array properties. Thereby, the $m$ set derived from (a,b) was used to normalize the dI(V)/dV spectra shown in Figs.1,~2,~5.

Noteworthily is to compare the shape of the Andreev dip in the spectra of arrays with various $m$. Fig.~6b shows the half-width $w$ of the main $n_L=1$ dips in the obtained Andreev spectra versus the corresponding number of junctions. In order to compare the data, it is reasonable to normalize $w$ by $\Delta_L(0)$ value or $k_BT_c$ in order to account even its minor variation from one contact point to another. The data spread modulating the general decrease could be caused by a broadening parameter $\Gamma$ which is, no doubt, different for each contact. Nonetheless, the covering tendency is, the larger $m$, the narrower are the Andreev dips in the dI(V)/dV spectrum (the tendency is shown in Fig.6b by a line).

A representative sample of the dip sharpening in the large-$m$ arrays is shown in the (g) panel. The fragments of dI(V)/dV curves with the first ($n_L=1$) gap feature are put together and shifted vertically for clarity; the position of the dip was normalized by a factor of $\Delta_L/e$. Such sharpening seems non surprising since in the arrays with large $m$, the contribution of the bulk rather than surface to the conductance becomes more significant.

Accordingly, on the one hand, the study of arrays is preferable since facilitates an observation of the bulk order parameter. On the other hand, measuring the dynamic conductance of the arrays with less $m$ is favourable as well in order to determine the gap magnitude and correct set of $m$ (see Fig.~6(c-f)). Whatever wide the Andreev feature is, this unaffects the dip position which directly scales with $\Delta$.


\begin{thebibliography}{}
\bibitem{remark_1} If so, the intergrain ScS junctions would have accidental and nonequivalent conductivity. As a result, the dI(V)/dV-spectrum is expected to show strongly smeared and irreproducible features, which drastically differs from that observed in our experiment \cite{EPL}.
\end{thebibliography}
\end{document}